# Universal mechanical exfoliation of large-area 2D crystals

Yuan Huang[1†], Yu-Hao Pan[2†], Rong Yang[1†], Li-Hong Bao[1], Lei Meng[1], Hai-Lan Luo[1], Yong-Qing Cai[1], Guo-Dong Liu[1], Wen-Juan Zhao[1], Zhang Zhou[1], Liang-Mei Wu[1], Zhi-Li Zhu[1], Ming Huang[3], Li-Wei Liu[4], Lei Liu[5], Peng Cheng[1], Ke-Hui Wu[1], Shi-Bing Tian[1], Chang-Zhi Gu[1], You-Guo Shi[1], Yan-Feng Guo[6], Zhi Gang Cheng[1,7,8], Jiang-Ping Hu[1,7,8], Lin Zhao[1,7,8], Eli Sutter[9], Peter Sutter[10*], Ye-Liang Wang[1,4], Wei Ji[2*], Xing-Jiang Zhou[1,7,8*], and Hong-Jun Gao[1,7*]

[1]Institute of Physics, Chinese Academy of Sciences, Beijing 100190, China

[2]Department of Physics and Beijing Key Laboratory of Optoelectronic Functional Materials & Micro-Nano Devices, Renmin University of China, Beijing 100872, China

[3]School of Materials Science and Engineering, Ulsan National Institute of Science and Technology (UNIST), Ulsan 44919, Republic of Korea

[4]School of Information and Electronics, MIIT Key Laboratory for Low-Dimensional Quantum Structure and Devices, Beijing Institute of Technology, Beijing, 100081, China

[5]College of Engineering, Peking University, Beijing 100871 China

[6]School of Physical Science and Technology, Shanghai Tech University, Shanghai 201210, China

[7]University of Chinese Academy of Sciences, Beijing, 100049, China

[8]Songshan Lake Materials Laboratory, Dongguan, 523808, China

[9]Department of Mechanical and Materials Engineering, University of Nebraska—Lincoln, Lincoln, Nebraska 68588, United States

[10]Department of Electrical and Computer Engineering, University of Nebraska—Lincoln, Lincoln, Nebraska 68588, United States

†These authors contributed equally to this work.
*Correspondence to: psutter@unl.edu (P.S.); wji@ruc.edu.cn (W.J.); xjzhou@iphy.ac.cn (X.J.Z.); hjgao@iphy.ac.cn (H.J.G.)

**Abstract**

Two-dimensional (2D) materials provide extraordinary opportunities for exploring phenomena arising in atomically thin crystals[1-4]. Beginning with the first isolation of graphene[5], mechanical exfoliation has been a key to provide high-quality 2D materials but despite improvements it is still limited in yield[6], lateral size and contamination. Here we introduce a contamination-free, one-step and universal Au-assisted mechanical exfoliation method and demonstrate its effectiveness by isolating 40 types of single-crystalline monolayers, including elemental 2D crystals, metal-dichalcogenides, magnets and superconductors. Most of them are of millimeter-size and high-quality, as shown by transfer-free measurements of electron microscopy, photo spectroscopies and electrical transport. Large suspended 2D crystals and heterojunctions were also prepared with high-yield. Enhanced adhesion between the crystals and the substrates enables such efficient exfoliation, for which we identify a common rule that underpins a universal route for producing large-area monolayers and thus supports studies of fundamental properties and potential application of 2D materials.

**Main**

Two-dimensional (2D) materials continue to reveal dimensionality-correlated quantum phenomena, such as 2D superconductivity, magnetism, topologically protected states, and quantum transport[1, 7-11]. Stacking 2D materials into van der Waals heterostructures leads to further emergent phenomena and derived device concepts[12, 13]. The further discovery and application of new properties depend on the development of synthesis strategies for 2D materials and heterostructures[6, 14-18]. Synthesis using crystal growth methods can now produce large (millimeter-scale) single crystals of some 2D materials, notably graphene and hexagonal boron nitride, but the scalable growth of high-quality crystals has remained challenging[18, 19], with many 2D materials and especially heterostructures proving difficult to realize by bottom-up approaches. Ion-intercalation and liquid exfoliation are used as top-down approaches but, as chemical methods, they often cause contamination of the isolated 2D surfaces[20, 21]. While some gold-assisted exfoliation methods were demonstrated in layered chalcogenides[22-24], those methods still bring unexpected contamination in samples prepared for electrical and optical measurements and thus reduce their performances when removing gold films with chemical solvents in additional steps. In light of this, a contamination-free, one-step and universal preparation strategy for large-area, high-quality monolayer materials is still lacking for both fundamental research and applications.

In the past 15 years, mechanical exfoliation has been a unique enabler of the exploration of new 2D materials. Most intrinsic properties of graphene, such as the quantum Hall effect[25], massless Dirac Fermions[26], and superconductivity[27], were mostly observed on exfoliated flakes but are either inaccessible or suppressed in samples prepared by other methods[17, 28]. While exfoliation often suffers from low yield and small sizes of the exfoliated 2D flakes[5], many layered materials are, however, yet to be exfoliated into monolayers by established exfoliation methods. Such challenge of exfoliation limits their utility for scalable production of 2D crystals and complicates further processing, *e.g.*, to fabricate heterostructures. These issues could be resolved by identifying suitable substrates that firmly adhere to 2D crystals without compromising their structure and properties, thus allowing the separation and transfer of the entire top sheet from a layered bulk crystal. Covalent-like quasi-bonding (CLQB), a recently uncovered non-covalent interaction with typical interaction energies of ~0.5 eV/unit cell[29-31], fits the requirements of the craving interaction between substrates and 2D layers. The intermediate interaction energy for CLQB is a balance of a reasonably large Pauli repulsion induced by interlayer wavefunction overlap and an enhanced dispersion attraction caused by more pronounced electron correlation in 2D layers with high polarizability.

Promising candidate substrates for CLQB with 2D crystals are materials whose Fermi level falls in a partially filled band with mostly *s*- or *p*-electrons to prevent disrupting the electronic structure of 2D layers, and which have highly polarizable electron densities to ensure a large dispersion attraction. Noble metals meet these criteria and are easily obtained as clean solid surfaces. Group 11 (IB) coinage metals, *i.e.*, Cu, Ag and Au, remain as potential candidates after ruling out Pt of high melting temperature, other group 8-10 (VIII) metals of too strong hybridization, Al of high activity to 2D layers and in air[32, 33] and closed-shell group 12 (IIB) metals (Zn, Cd, Hg). Among those three, Au interacts strongly with group 16 (VIA) chalcogens (S, Se, Te) and 17 (VIIA) halogens (Cl, Br, I), which terminate surfaces in most 2D materials. Together with its low chemical reactivity and air stability, Au appears promising for high-yield exfoliation of many 2D materials, which is also evidenced by three previous attempts[22-24].

Density functional theory (DFT) calculations were employed to substantiate these arguments by comparing the interlayer binding energies of a large set of layered crystals with their adhesion energies to the Au (111) surface. A total of 58 layered materials, including 4 non-metallic elemental layers and 54 compounds comprised of metal and non-metal elements were considered in our calculations (see Fig. 1a). They belong to 18 space groups covering square, hexagonal, rectangular and other lattices (Fig. 1b). The surfaces of all considered compound layers are usually terminated with group 16 (VIA) or 17 (VIIA) elements, *e.g.*, S, Se, Te, Cl, Br and I, with the exception of $W_2N_3$. These atoms, together with group 15 (VA) elements, are expected to have substantial interactions with Au substrates, which is verified by our differential charge density (DCD) plots.

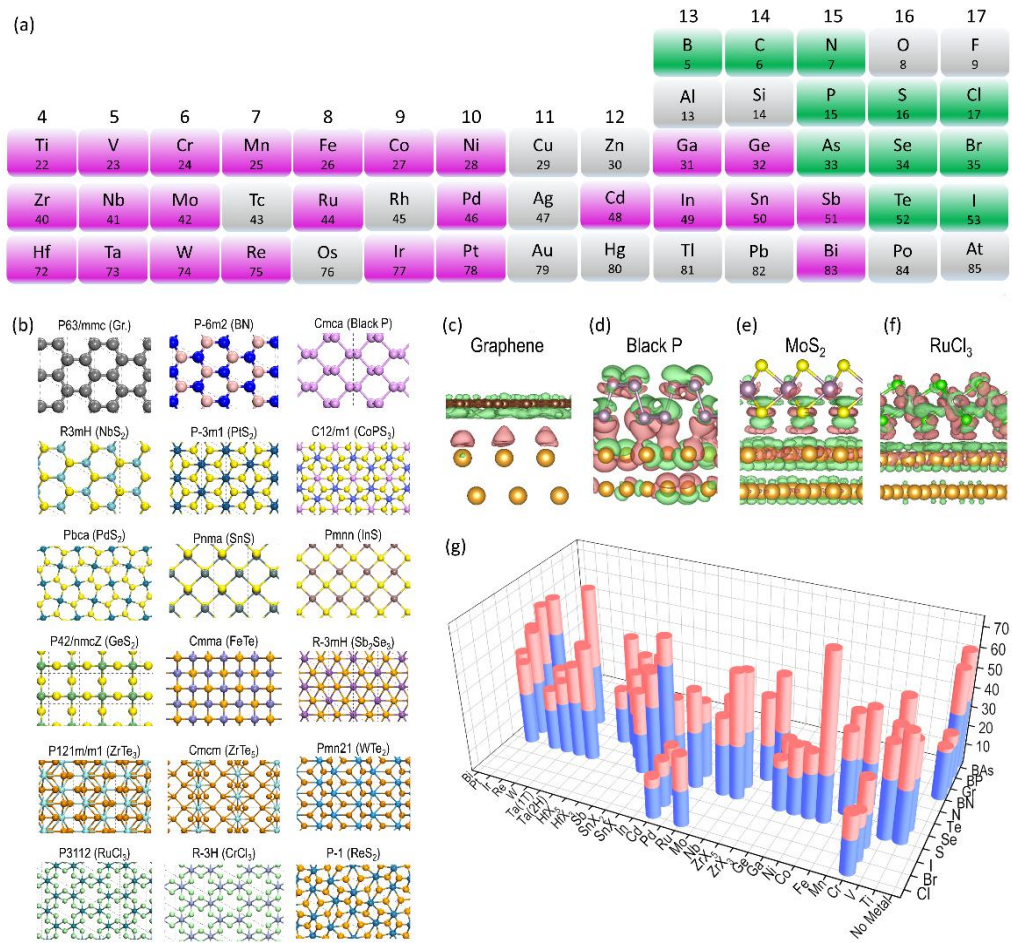

**Fig. 1. DFT calculated interlayer binding energies of 2D materials and adsorption energies on Au (111) surfaces. (a)** Part of the periodic table, showing the elements involved in most 2D materials between groups 4 (IVB) and 17 (VIIA). **(b)** Eighteen space groups and typical structural configurations (top views) of the 2D materials. **(c-f)** DCD of four Au (111)/2D crystal interfaces with (non-metallic) terminating atoms between groups 14 (IVA) and 17 (VIIA). Isosurface values of these DCD plots are $5\times10^{-4}$ $e$/Bohr$^3$ (graphene), $1\times10^{-2}$ $e$/Bohr$^3$ (BP), and $1\times10^{-3}$ $e$/Bohr$^3$ (MoS$_2$, RuCl$_3$), respectively. **(g)** Bar graph comparing the interlayer binding energies of 2D materials (blue cylinders) with their adsorption energies on Au (111) (red cylinders). The visible red cylinders represent the difference between the Au/2D crystal interaction and the interlayer interaction.

Figs. 1c-f show the DCDs of graphene, black phosphorus (BP), MoS$_2$, and RuCl$_3$ monolayers adsorbed on Au (111), representing the interactions of Au with group 14 (IVA) to 17 (VIIA) atoms, respectively. The adhesion induced charge redistribution of graphene differs from those of the other three layers. While Au only introduces charge dipoles at the interface to graphene, significant covalent characteristics, *i.e.*, charge reduction near the interfacial atoms and charge accumulation between them, were observable at the P/Au, S/Au and Cl/Au interfaces. The difference in charge redistribution is reflected in the smaller adhesion energy of graphene/Au (28 meV/Å$^2$; 0.15 eV/unit cell) compared with those of the other three interfaces (56, 40 and 36 meV/Å$^2$; 0.80, 0.35 and 1.11 eV/unit cell). The clearly covalent nature of the S/Au interface is consistent with previous reports[22-24] and confirms our expectation. Our results of DCD and electronic band structures (Fig. S1), suggest the existence of CLQB at the S/Au, P/Au and Cl/Au interfaces, which is confirmed by comparing the interlayer (0.23, 0.48 and 0.57 eV/unit cell) and 2D crystal/Au (0.35, 0.80 and 1.11 eV/unit cell) binding energies. Fig. 1g and Table S1 show the comparison of these energies for all 57 considered 2D crystals, where the 2D crystal/Au binding is invariably stronger than the corresponding interlayer binding. These results support the concept that the 2D crystal/Au interaction should be sufficient to overcome the interlayer attraction and facilitate exfoliating monolayers from a broad range of layered crystals. Here, we define a ratio $R_{LA/IL}$ as layer-Au over interlayer adhesion energies. Possible exceptions are those 2D materials whose $R_{LA/IL}$ values, while greater than 1, are substantially smaller than usual $R_{LA/IL}$ values (>1.3). Here, BN (1.07), GeS$_2$ (1.17) and graphene (1.24) are some examples.

To test these theoretical predictions, we implemented the Au-assisted exfoliation of 2D materials as shown in Fig. 2a. Firstly, a thin layer of Au is deposited onto a substrate covered with a thin Ti or Cr adhesion layer. Then, a freshly cleaved layered crystal is brought in contact with the Au layer. Adhesive tape is placed on the outward side of the crystal, and gentle pressure is applied to establish a good layered crystal/Au contact. Peeling off the tape removes the major portion of the crystal, leaving one or few large-area monolayer flakes on the Au surface. Limited only by the size of available bulk crystals, these monolayer flakes are usually macroscopic in size (millimeters; see Methods for details).

Optical microscopy was used to examine the dimensions and uniformity of the exfoliated 2D crystals. Fig. 2b shows an image of exfoliated MoS$_2$ monolayers reaching lateral dimensions close to 1 cm on a SiO$_2$/Si substrate covered with Au (2 nm)/Ti (2 nm). We also extended the base substrate from the SiO$_2$/Si substrate to transparent (quartz, sapphire; Fig. 2c) and flexible plastic supports (Fig. 2d). The transparency persists even for thicker (~10 nm) Au/Ti layers although light transmission slightly decreases. This method can also be applied to CVD-grown wafer-scale transition-metal dichalcogenides (TMDCs) materials, such as $MoS_2$ (Fig. 2e).

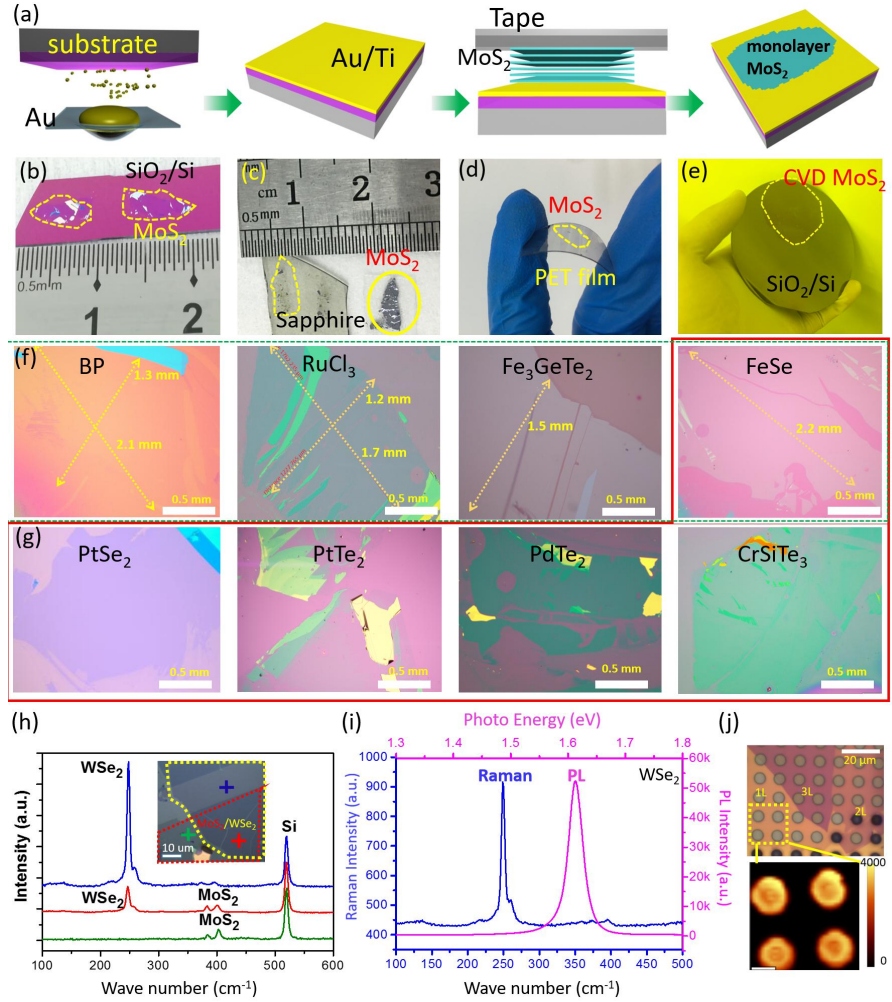

**Fig. 2. Mechanical exfoliation of different monolayer materials with macroscopic size. (a)** Schematic of the exfoliation process. **(b-d)** Optical images of exfoliated $MoS_2$ on $SiO_2$/Si, sapphire, and plastic film. **(e)** 2-inch CVD-grown monolayer $MoS_2$ film transferred onto a 4 inch $SiO_2$/Si substrate. **(f-g)** Optical images of large exfoliated 2D crystals: BP, FeSe, $Fe_3GeTe_2$, $RuCl_3$, $PtSe_2$, $PtTe_2$, $PdTe_2$ and $CrSiTe_3$. Those exfoliated monolayers highlighted in the red box are, so far, not accessible using other mechanical exfoliate method. **(h)** Optical image and Raman spectra of a $MoS_2/WSe_2$ heterostructure. **(i)** Raman and photoluminescence (PL) spectra of suspended monolayer $WSe_2$. **(j)** Optical image of suspended $WSe_2$ with different thicknesses (1L to 3L) and a PL intensity map of the suspended monolayer.

X-ray photoelectron spectroscopy (XPS) was employed to further investigate the interaction between $MoS_2$ and Au. Fig. S2d shows an XPS spectrum of exfoliated

MoS$_2$ near the Mo 3d region. Peaks centered at 226.5 eV, 229 eV, and 232 eV result from S 2*s*, Mo 3*d$_{5/2}$* and Mo 3*d$_{3/2}$* photoelectrons, respectively. There are no appreciable changes in terms of shape, binding energy, and width of the XPS peaks compared to those of bulk MoS$_2$. Hence, the nearly unchanged XPS spectra confirm CLQB rather than covalent bonding between MoS$_2$ and the Au substrate. Figs. S3 and S4 show Raman, photoluminescence (PL) and ARPES of a typical exfoliated MoS$_2$ monolayer. Sharp $E_{2g}$ and $A_{1g}$ Raman peaks at 386 and 406 cm$^{-1}$, respectively, confirm the high quality of the MoS$_2$ monolayer[34]. The pronounced A-exciton PL peak at 1.83 eV indicates the exfoliated MoS$_2$ on Au is still a direct-band gap semiconductor[35]. Metal substrates usually quench the PL intensity of monolayer MoS$_2$[23, 24], however, our PL signal remains strong because of the thickness-tunable conductivity of our metal substrates, as elucidated later.

We applied the Au-assisted exfoliation method to other 2D crystals and have obtained a library of 40 large-area single-crystal monolayers, as shown in Fig. 2f-2g and Fig. S5. Besides transition metal dichalcogenides, the library contains metal monochalcogenides (*e.g.*, GaS), black phosphorus, black arsenic, metal trichlorides (*e.g.*, RuCl$_3$) and magnetic compounds (*e.g.*, Fe$_3$GeTe$_2$). It is rather striking that some monolayers, i.e. FeSe, PdTe$_2$, CrSiTe$_3$, become accessible by our exfoliation method for the first time. This method is, as we expected according to the smaller $R_{LA/IL}$ values (1.24 and 1.07), less effective for exfoliating graphene and h-BN monolayers which are accessible by chemical vapor deposition. The exfoliated monolayer samples show high quality, as characterized by Raman and atomic force microscopy (Figs. S6 and S7). Reactive samples were exfoliated in a glove box due to their stability issues in air.

Our method also promotes preparation of van der Waals heterostructures and suspended 2D materials at human visible size scales. Fig. 2h shows a typical monolayer MoS$_2$/WSe$_2$ heterostructure prepared using this method. Raman spectra (Fig. 2h) show the characteristic vibrational modes of both the MoS$_2$ and WSe$_2$ layers. Given the exceeding $R_{LA/IL}$ values over 1.3, patterned Au thin-films on substrates with holes, are also, most likely, able to exfoliate 2D crystals and thus to fabricate suspended monolayers, which is of paramount importance on studying intrinsic properties of 2D layers[36, 37]. We show an example with suspended 1L-3L WSe$_2$ in (Fig. 2i), which can reach 90% coverage over at least tens of micrometers (Fig. S8). The suspended monolayer film is detached from multilayer instead of transferring monolayer by organic films, which totally avoid polymer contamination. In comparison with supported samples on SiO$_2$, the suspended WSe$_2$ (Fig. 2i) shows enhanced PL intensity (16 times) and sharper PL peak (full width at half maximum (FWHM): 34 meV, compared with 64 meV for supported WSe$_2$) as shown in Fig. S9. Since PL can be fully quenched on thicker metal film while well maintained on suspended area, therefore, we realized patterning of PL even on one monolayer flake (Fig. 2j).

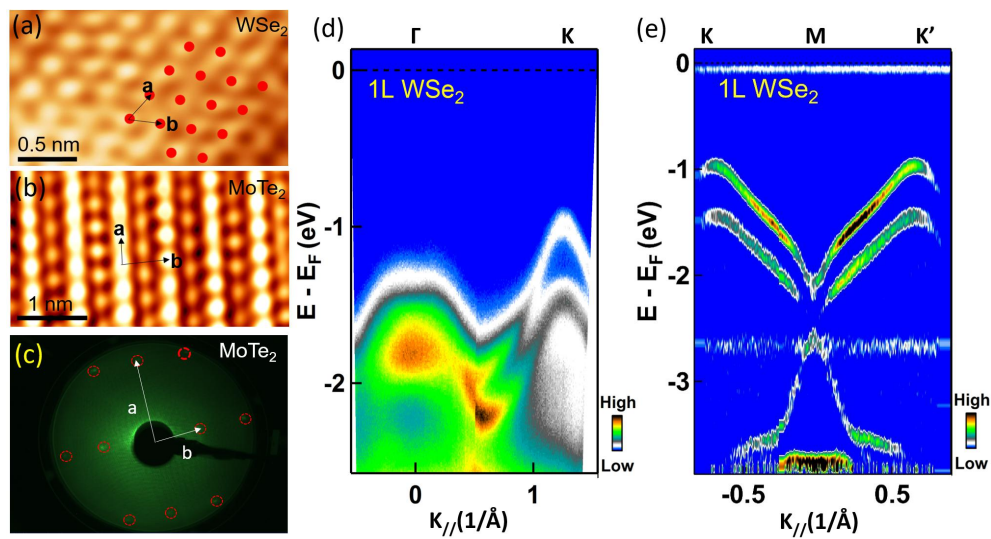

**Fig. 3. STM and ARPES measurements of 2D materials exfoliated onto conductive Au/Ti adhesion layers. (a)** and **(b)** STM images of monolayer WSe$_2$ and T$_d$-MoTe$_2$, respectively. **(c)** LEED pattern of monolayer T$_d$-MoTe$_2$. **(d)**, **(e)** Band structure of monolayer WSe$_2$. **(d)** Original ARPES band structure of monolayer WSe$_2$ ($hv$ = 21.2 eV) along Γ-K high symmetry line. The valence band maximum (VBM) is positioned at K instead of Γ, which is an important signature of monolayer WSe$_2$. **(e)** Second-derivative spectra of band dispersion along K-M-K', showing clear spin-orbital coupling (SOC) induced spin-splitting bands.

High-quality macroscopic monolayers have practical advantages, for instance in establishing the lattice structure and electronic band structure of unexplored 2D materials or van der Waals stacks by scanning tunneling microscopy (STM) and ARPES. The Au-coated support facilitates such electron-based spectroscopy by eliminating charging effects associated with insulating (*e.g.*, SiO$_2$) substrates while preserving the intrinsic electronic band structures. Figs. 3a and 3b illustrate atomic-resolution STM images for as-exfoliated WSe$_2$ and T$_d$-MoTe$_2$ monolayers, which are challenging to image on insulating substrates due to charging effects. Low-energy electron diffraction with millimeter incident electron beam size shows a single-phase diffraction pattern for MoTe$_2$ (Fig. 3c), indicating that it is a single-crystal at the millimeter scale. Fig. 3d displays an ARPES map of the low-energy electronic structure of the WSe$_2$ monolayer, showing clear and sharp bands. The valence band features a single flat band around Γ and a large band splitting near K. Along the Γ-K line, one single band starts to split into two spin-resolved bands at $k \approx \frac{1}{3}\Delta k^{K,\Gamma}$, and the valence band maximum at K sits at ~0.6 eV higher than that at Γ. Fig. 3e displays the symmetric band splitting spectra along K-M-K' arising from strong spin-orbit coupling mainly at the W site in the WSe$_2$ lattice[38]. These features constitute the critical signatures of band dispersion in monolayer TMDCs. Here it deserves an emphasis on the big advantage of large area of monolayer TMDCs, which make it quite feasible and easy to accurately measure the band structure by using standard ARPES technique[39].

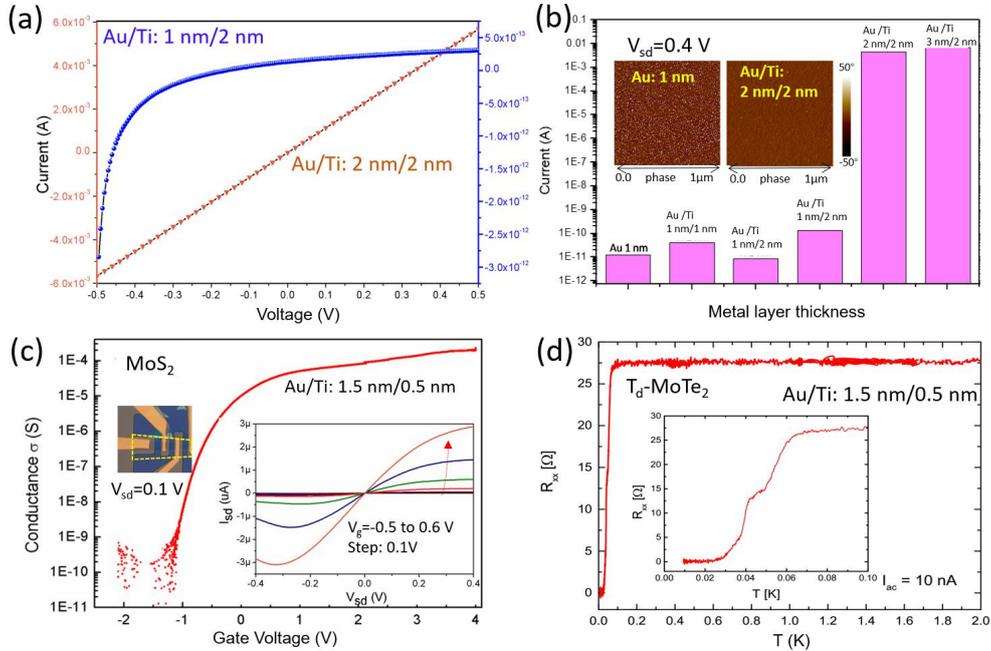

**Fig. 4. Electrical measurements of metal adhesion layers and of 2D materials exfoliated onto non-conductive metal films. (a)** Electrical transfer curves of typical Au/Ti adhesion layers. **(b)** Two-terminal resistance of Au/Ti layers with different nominal thickness. The inset shows AFM phase maps of two metal layers. **(c)** Gate voltage-conductance transfer characteristics of a top-gated MoS$_2$

FET on SiO$_2$/Si with Au (1.5 nm)/Ti (0.5 nm) adhesion layer (T = 220 K, source−drain bias V$_{sd}$= 0.1 V). Left inset: Optical image of the FET device with windows for the ionic liquid top gate. Right inset: low-bias source-drain current-voltage characteristics for gate voltage -0.5 - 0.6 V. **(d)** Temperature-dependent resistance of a T$_d$-MoTe$_2$ flake exfoliated onto SiO$_2$/Si with a 2 nm metal adhesion layer.

Even some previous studies claimed to exfoliate large area chalcogenides layers using gold films[22-24], however, further characterizations including optical and electrical measurements are usually achieved by an additional transfer process onto insulating substrate. Nevertheless, we show here that the conductivity of Au/Ti adhesion layer can be drastically tuned by controlling its nominal thickness, which demonstrate that both optical measurements and device fabrication can be realized on this one-step exfoliated samples. Fig. 4a and 4b show that the Au/Ti films become insulating (*i.e.*, electrically discontinuous) if the combined thickness of Ti/Au decreases to 3 nm or below[40]. Our exfoliation method is not apparently limited by the Au thickness, therefore, those large-area 2D crystals are expected to be exfoliated by either conducting or non-conducting Au coated substrates.

Fig. S3c shows a MoS$_2$ flake exfoliated by an electrically discontinuous (0.5 nm Ti, 1.5 nm Au) adhesion layer on a SiO$_2$/Si substrate (Fig. 4b inset and Fig. S10), in which almost no lateral channel is available to carry current flow through the substrate. Although some prior studies used Au films to enhance exfoliation of MoS$_2$, the PL intensity of their MoS$_2$ samples was largely quenched[22, 24]. By contrast, our MoS$_2$ monolayer exfoliated using electrically discontinuous metal adhesion layers show intense PL signals (see Fig. S3). Such electrically discontinuous layers also allow to fabrication of electronic devices directly from the as-exfoliated 2D monolayers. Fig. 4c shows the trans-conductance curve of a prototype device, a field-effect transistor (FET) directly built on an as-exfoliated monolayer MoS$_2$ channel on Au(1.5 nm)/Ti(0.5 nm)/SiO$_2$/Si. Fig. S11 shows the device layout. The device, controlled by an ionic-liquid top gate, shows a high on-off current ratio (>10$^6$ at *T* = 220 K), comparable to usual MoS$_2$ FETs directly fabricated on SiO$_2$/Si substrates[7, 41]. A sub-threshold swing (SS) of 100 mV/dec. was derived, close to the best values reported in the literature[7, 42, 43], ranging from 74 mV/dec. to 410 mV/dec.

Both results show good performance of the FET directly fabricated on the ultrathin metal adhesion layer and its potential for further improvements. We extended the individual FET to an FET array directly built on an exfoliated centimeter-scale single-crystal MoS$_2$ flake using UV lithography (Fig. S12), indicating great potential of our new exfoliation method for fabrication of integrated circuit.

Since electrical current is prone to flow through superconducting regions, the metal adhesion layer does not suppress the transition of an exfoliated layer from its normal state to its superconducting state. Fig. 4d shows a T$_d$-MoTe$_2$ device, directly fabricated on a 2 nm metal layer, undergoes a metal-superconductor transition at 70 mK, and reaches the zero-resistance at 30 mK. A second onset, observed at 50 mK, is, most likely, a result of quantum fluctuation in 2D crystals. A previous study of bulk T$_d$-MoTe$_2$ showed an onset of superconductivity at 250 mK and zero resistance at a critical temperature $T_c$=100 mK[44]. The difference between bulk and monolayer's $T_c$ values may be primarily relevant with reduced dimensionality[45].

Our combined results show that exfoliation assisted by an Au adhesion layer with covalent-like quasi bonding to a layered crystal provides access to an unprecedentedly broad spectrum of large-area monolayer materials. This method is rather unique, especially for layered crystals that are difficult to exfoliate using conventional methods. The versatility of this approach is demonstrated here by using Au adhesion layers for exfoliation of large 2D sheets from 40 layered materials. The efficient transfer of most 2D crystals is rationalized by calculations that indicate interaction energies to Au exceeding the interlayer energy for most layered bulk crystals, graphene and hexagonal boron nitride being notable exceptions. Characterization of the large-area exfoliated monolayers flakes demonstrates that the flakes are of high quality. For research on atomically thin materials, the approach demonstrated here has immediate implications. The availability of macroscopic (millimeter scale) 2D materials can support the exploration of the properties of new families of ultrathin semiconductors, metals, superconductors, topological insulators, ferroelectrics, etc., as well as engineered van der Waals heterostructures. For applications of 2D materials, an efficient large-scale layer transfer method could force a paradigm shift. So far, exfoliation from bulk crystals has not been deemed technologically scalable. But once exfoliation becomes so consistent that the size of the resulting 2D layers is limited only by the dimensions and crystallinity of the source crystal, the focus of application-driven materials research may shift toward optimizing the growth of high-quality layered bulk crystals. Ironically, the fabrication of 2D materials for applications would then follow the well-established and highly successful example of silicon technology, where the extraction of wafers from large, high-quality single crystals has long been key to achieving the yields and reliability required for industrial applications.

## Materials and Methods

***DFT calculations***: DFT calculations were performed using the generalized gradient approximation for the exchange-correlation potential, the projector augmented wave method [46, 47], and a plane-wave basis set as implemented in the Vienna *ab-initio* simulation package (VASP) [48]. The energy cutoff for plane wave was set to 700 and 500 eV for variable volume structural relaxation of pure 2D materials and invariant volume structural relaxation of these materials on Au (111) surface. Dispersion correction was made at the van der Waals density functional (vdW-DF) level, with the optB86b functional for the exchange potential[49]. Seven Au (111) layers, separated by a 15 Å vacuum surface slab, were employed to model the surface. The four bottom layers were kept fixed and all other atoms were fully relaxed until the residual force per atom was less than 0.04 eV·Å$^{-1}$ during the relaxations of 2D materials on the Au (111) surfaces and less than 0.01 eV·Å$^{-1}$ during all the other structural relaxations. The lattice constancies of Au (111) surface model were changed to match those of 2D materials for keeping electronic properties of 2D materials unchanged by strains or stresses. The lattice mismatches between 2D materials and Au (111) surface were kept lower than 4.5%.

***Gold-enhanced mechanical exfoliation***: The metal layer deposition was completed in an electron evaporation system (Peva-600E). An adhesion metal layer (Ti or Cr) was first evaporated on Si substrate (with 300 nm SiO$_2$ film), after that Au film was deposited on the substrate. The thickness of Ti (or Cr) and Au can be well controlled by the evaporation rate (0.5 Å/s). After depositing metal layers on Si wafer, a fresh surface of layered crystal was cleaved from tape and put it onto the substrate. By pressing the tape vertically for about 1 min, the tape can be removed from substrate. Large area monolayer flakes can be easily observed by optical microscope or even by eyes. Most of the time, the size of monolayer flakes is limited by the size of bulk crystal.

***Suspended samples preparation***: Si wafer with 300 nm SiO$_2$ was patterned by UV-lithography, after that the hole array structures were prepared by reactive-ion etching. The diameter and depth of each hole is 5 μm and 10 μm. The metal layers (Au/Ti: 2 nm/2 nm) deposited on the Si substrate with hole array before exfoliating layered materials on it. Large area suspended 2D materials can be exfoliated on the hole array substrate.

***Characterization and measurement***:

**The Raman and PL measurements** were performed on a JY Horiba HR800 system with a wavelength of 532 nm and power at 0.6 mW. Fig. S3 shows the Raman, and PL results of a typical exfoliated $MoS_2$ film. Raman peaks of $E^1_{2g}$ and $A_{1g}$ are at 386 and 406 cm$^{-1}$, respectively. From the Raman result, we can determine the following: first, the space between the two peaks ($\Delta$) is ~20 cm$^{-1}$, suggesting that the as-exfoliated material is monolayer; second, the Raman peaks show no split, suggesting that the crystal quality of the as-grown $MoS_2$ film is good. The PL peak of A exciton is at 1.83 eV. These features are in good agreement with the data seen previously for monolayer $MoS_2$.

Fig. S6 presents the representative Raman spectra for monolayer and few-layer BP and α-$RuCl_3$ samples excited by 2.33 eV radiation in vacuum environments. The laser power on the sample during Raman measurement was kept below 100 μW in order to avoid sample damage and excessive heating. The silicon Raman mode at 520.7 cm$^{-1}$ was used for calibration prior to measurements and as an internal frequency reference.

In close analogy to bulk, three typical Raman peaks are resolved at 360, 436, and 465 cm$^{-1}$ for monolayer and few-layer BP samples. It can be seen that the phonon intensity decreases with layer numbers. The evolution of Raman intensity versus the number of layers is attributed to the multilayer interference occurring for both the incident light and the emitted Raman radiation, being akin to the case of graphene[50] and $MoS_2$[51, 52].

As for monolayer and few-layer α-$RuCl_3$ samples, five strong and sharp phonon modes are resolved at 117, 164, 270, 296 and 312 cm$^{-1}$. It should be noted that the two lowest energy phonons in α-$RuCl_3$ show asymmetric Fano line shape stemmed from the coupling between the discrete optical phonons and the magnetic scattering, as we have discussed before[53].

For all thicknesses, it can be seen that the energies of all phonons are independent on the number of layers, as indicated by the dashed vertical lines in Fig. S6. This is in marked contrast to TMDC[33] and indicates that the van der Waals interlayer interactions in BP and α-$RuCl_3$ are extremely weak and have tiny effects on phonon energy.

**The atomic force microscope (AFM) scanning** (Veeco Multimode III) was used to check the thickness and surface morphology of those monolayer samples. Fig. S7 shows the AFM images of atomically thin $MoS_2$ and BP flakes. Judging from the morphology of the freshly exfoliated samples at the nanometer scale, there is no evidence of structural irregularity or bubbles on the surfaces. The height profiles taken along the blue lines in AFM images are depicted below. It is noted that the height of monolayer $MoS_2$/BP on $SiO_2$/Si substrate is about ~0.65 nm/~0.6 nm, a little larger than the theoretical thickness. The deviation implies that there are some absorbents at the interface between the $MoS_2$/BP and $SiO_2$/Si substrate[51, 54].

**The X-ray photoelectron spectroscopy** (XPS, Thermo Scientific ESCALAB 250 Xi) was performed with Al Kα X-rays (hυ = 1486.6 eV) in an analysis chamber that had a base pressure < $3 \times 10^{-9}$ Torr. Core spectra were recorded using a 50 eV constant pass energy (PE) in 50-100 μm small area lens mode (i.e., aperture selected area). The XPS peaks were calibrated using the adventitious carbon C1s peak position (284.8 eV). XPS analysis was employed to obtain the core-level XPS spectra for the comparison of exfoliated single layer $MoS_2$ and bulk $MoS_2$ flakes (Fig. S2). The high-resolution spectra and the fitting curves of Au 4f peaks in Fig. S4a, b do not show noticeable difference of the binding energies but with a slightly decreased intensity for the Au spectra under $MoS_2$ flake due to the screening effect. While the core-level Mo 3d and S 2s spectra of the exfoliated $MoS_2$ show a small redshift (around 0.2 eV) when compared with the bulk $MoS_2$ bulk flakes, suggesting a possible charge transfer between the single layer $MoS_2$ flake and the underlying Au film[22, 55]. In addition, the consistence of the shape and peak width of the Mo 3d spectra of the exfoliated single layer $MoS_2$ with the bulk $MoS_2$ flake indicates that the single layer $MoS_2$ is clean and retains its chemical identity after exfoliation. This XPS result demonstrates the important role of Au for the successful exfoliation of large single layer $MoS_2$ flake.

**The scanning tunneling microscope (STM) measurement** was performed using a custom built, low-temperature, and UHV STM system at 300 K. A chemically etched W STM tip was cleaned and calibrated against a gold (111) single crystal prior to the measurements.

**High resolution angle-resolved photoemission spectroscopy (ARPES) measurements** were carried out on our lab system equipped with a Scienta R4000 electron energy analyzer[56]. We use Helium discharge lamp as the light source which can provide photon energies of $h\upsilon$ = 21.218 eV (Helium I). The energy resolution was set at 10~20 meV for band structure measurements (Fig. 3). The angular resolution is ~0.3 degree. The Fermi level is referenced by measuring on a clean polycrystalline gold that is electrically connected to the sample. The samples were measured in vacuum with a base pressure better than $5 \times 10^{-11}$ Torr.

**The electrical characteristic measurements** were carried out in the probe station with the semiconductor parameter analyzers (Agilent 4156C and B1500) and oscilloscope.

**Online Content**

Methods, along with any additional Extended Data display items and Source Data, are available in the online version of the paper; references unique to these sections appear only in the online paper.

**Acknowledgments**

This work is supported by the National Key Research and Development Program of China (Grant No. 2019YFA0308000, 2018YFA0305800, 2018YFE0202700), the Youth Innovation Promotion Association of CAS (2019007, 2018013), the National Natural Science Foundation of China (Grant No. 11874405, 11622437, 61674171 and 11974422), the National Basic Research Program of China (Grant No. 2015CB921300), and the Strategic Priority Research Program (B) of the Chinese Academy of Sciences (Grant No. XDB07020300, XDB30000000), the Research Program of Beijing Academy of Quantum Information Sciences (Grant No. Y18G06). P.S. and E.S. acknowledge support by the U.S. Department of Energy, Office of Science, Basic Energy Sciences, under Award No. DE-SC0016343. Calculations were performed at the Physics Lab of High-Performance Computing of Renmin University of China and Shanghai Supercomputer Center.

**Author Contributions**

P.S., W.J., H.J.G. and X.J.Z. are equally responsible for supervising the discovery. Y.H. and R.Y. conceived the project. Y.H.P., J.P.H. and J.W. performed the DFT calculations. Y.H., H.L.L and L.L. prepared all the mechanical exfoliation samples. Y.H. and R.Y. performed the Raman, PL and AFM measurements. M.H., J.W., P.S. and E.S. performed XPS measurement. Y.Q.C., G.D.L., L.Z. and W.J.Z. performed the ARPES measurement. Z.L.Z., P.C., K.H.W., L.M., Z.Z. and Y.L.W performed the STM and LEED measurement. Y.G.S. and Y.F.G. prepared bulk layered crystals. S.B.T., C.Z.G., Z.G.C., L.M.W. and L.H.B. fabricated the transistors and performed the electrical measurements. Y.H., R.Y., J.P.H., P.S., and W.J. analyzed data, wrote the manuscript and all authors discussed and commented on it. These authors contributed equally: Y.H, Y.H.P and R.Y.

**Competing interests**

Three Chinese patents were filed (DIC19110040, DIC19110039 and DIC19110044) by the Institute of Physics, Chinese Academy of Sciences, along with their researchers (Y.H., H.L.L and X.J.Z).

**References**

1. He, S.L. et al. Phase diagram and electronic indication of high-temperature superconductivity at 65 K in single-layer FeSe films. *Nature Materials* **12**, 605-610 (2013).
2. Chen, G.R. et al. Evidence of a gate-tunable Mott insulator in a trilayer graphene moire superlattice. *Nature Physics* **15**, 237-241 (2019).
3. Ezawa, M. Valley-Polarized Metals and Quantum Anomalous Hall Effect in Silicene. *Physical Review Letters* **109**, 055502 (2012).
4. Jiang, S.W., Li, L.Z., Wang, Z.F., Mak, K.F. & Shan, J. Controlling magnetism in 2D CrI3 by electrostatic doping. *Nature Nanotechnology* **13**, 549-553 (2018).
5. Novoselov, K.S. et al. Electric field effect in atomically thin carbon films. *Science* **306**, 666-669 (2004).
6. Huang, Y. et al. Reliable Exfoliation of Large-Area High-Quality Flakes of Graphene and Other Two-Dimensional Materials. *Acs Nano* **9**, 10612-10620 (2015).
7. Radisavljevic, B., Radenovic, A., Brivio, J., Giacometti, V. & Kis, A. Single-layer MoS2 transistors. *Nature Nanotechnology* **6**, 147-150 (2011).
8. McGuire, M.A., Dixit, H., Cooper, V.R. & Sales, B.C. Coupling of Crystal Structure and Magnetism in the Layered, Ferromagnetic Insulator CrI3. *Chemistry of Materials* **27**, 612-620 (2015).
9. Jiang, D. et al. High-T-c superconductivity in ultrathin $Bi_2Sr_2CaCu_2O_{8+x}$ down to half-unit-cell thickness by protection with graphene. *Nature Communications* **5** (2014).
10. Tang, S.J. et al. Quantum spin Hall state in monolayer 1T '-$WTe_2$. *Nature Physics* **13**, 683-687 (2017).
11. Zhang, Y. et al. Superconducting Gap Anisotropy in Monolayer FeSe Thin Film. *Physical Review Letters* **117**, 117001 (2016).
12. Novoselov, K.S., Mishchenko, A., Carvalho, A. & Neto, A.H.C. 2D materials and van der Waals heterostructures. *Science* **353**, 9439 (2016).
13. Liu, Y. et al. Van der Waals heterostructures and devices. *Nature Reviews Materials* **1** (2016).
14. Wang, C. et al. Monolayer atomic crystal molecular superlattices. *Nature* **555**, 231-236 (2018).
15. Li, X.S., Cai, W.W., Colombo, L. & Ruoff, R.S. Evolution of Graphene Growth on Ni and Cu by Carbon Isotope Labeling. *Nano Letters* **9**, 4268-4272 (2009).
16. Moon, I.K., Lee, J., Ruoff, R.S. & Lee, H. Reduced graphene oxide by chemical graphitization. *Nature Communications* **1**, 73 (2010).
17. Sutter, P.W., Flege, J.I. & Sutter, E.A. Epitaxial graphene on ruthenium. *Nature Materials* **7**, 406-411 (2008).
18. Meng, L. et al. Buckled Silicene Formation on Ir(111). *Nano Letters* **13**, 685-690 (2013).
19. Liu, H., Du, Y.C., Deng, Y.X. & Ye, P.D. Semiconducting black phosphorus: synthesis, transport properties and electronic applications. *Chemical Society Reviews* **44**, 2732-2743 (2015).
20. Coleman, J.N. et al. Two-Dimensional Nanosheets Produced by Liquid Exfoliation of Layered Materials. *Science* **331**, 568-571 (2011).
21. Fan, X.B. et al. Fast and Efficient Preparation of Exfoliated 2H $MoS_2$ Nanosheets by Sonication-Assisted Lithium Intercalation and Infrared Laser-Induced 1T to 2H Phase Reversion. *Nano Letters* **15**, 5956-5960 (2015).
22. Desai, S.B. et al. Gold-Mediated Exfoliation of Ultralarge Optoelectronically-Perfect

Monolayers. *Advanced Materials* **28**, 4053-4058 (2016).

23. Velicky, M. et al. Mechanism of Gold-Assisted Exfoliation of Centimeter-Sized Transition-Metal Dichalcogenide Monolayers. *ACS Nano* **12**, 10463-10472 (2018).

24. Magda, G.Z. et al. Exfoliation of large-area transition metal chalcogenide single layers. *Scientific Reports* **5** (2015).

25. Zhang, Y.B., Tan, Y.W., Stormer, H.L. & Kim, P. Experimental observation of the quantum Hall effect and Berry's phase in graphene. *Nature* **438**, 201-204 (2005).

26. Novoselov, K.S. et al. Two-dimensional gas of massless Dirac fermions in graphene. *Nature* **438**, 197-200 (2005).

27. Cao, Y. et al. Unconventional superconductivity in magic-angle graphene superlattices. *Nature* **556**, 43-50 (2018).

28. Hao, Y.F. et al. Oxygen-activated growth and bandgap tunability of large single-crystal bilayer graphene. *Nature Nanotechnology* **11**, 426-431 (2016).

29. Qiao, J.S., Kong, X.H., Hu, Z.X., Yang, F. & Ji, W. High-mobility transport anisotropy and linear dichroism in few-layer black phosphorus. *Nature Communications* **5**, 4475 (2014).

30. Hu, Z.-X., Kong, X., Qiao, J., Normand, B. & Ji, W. Interlayer electronic hybridization leads to exceptional thickness-dependent vibrational properties in few-layer black phosphorus. *Nanoscale* **8**, 2740-2750 (2016).

31. Qiao, J. et al. Few-layer Tellurium: one-dimensional-like layered elementary semiconductor with striking physical properties. *Science Bulletin* **63**, 159-168 (2018).

32. Li, W. et al. Experimental realization of honeycomb borophene. *Science Bulletin* **63**, 282-286 (2018).

33. Deng, Y. et al. Gate-tunable room-temperature ferromagnetism in two-dimensional $Fe_3GeTe_2$. *Nature* **563**, 94-99 (2018).

34. Li, H. et al. From Bulk to Monolayer $MoS_2$: Evolution of Raman Scattering. *Advanced Functional Materials* **22**, 1385-1390 (2012).

35. Mak, K.F., Lee, C., Hone, J., Shan, J. & Heinz, T.F. Atomically Thin $MoS_2$: A New Direct-Gap Semiconductor. *Physical Review Letters* **105**, 136805 (2010).

36. Du, X., Skachko, I., Barker, A. & Andrei, E.Y. Approaching ballistic transport in suspended graphene. *Nature Nanotechnology* **3**, 491-495 (2008).

37. Lee, C., Wei, X.D., Kysar, J.W. & Hone, J. Measurement of the elastic properties and intrinsic strength of monolayer graphene. *Science* **321**, 385-388 (2008).

38. Zhang, Y. et al. Electronic Structure, Surface Doping, and Optical Response in Epitaxial $WSe_2$ Thin Films. *Nano Letters* **16**, 2485-2491 (2016).

39. Zhang, H.Y. et al. Resolving Deep Quantum-Well States in Atomically Thin $2H-MoTe_2$ Flakes by Nanospot Angle-Resolved Photoemission Spectroscopy. *Nano Letters* **18**, 4664-4668 (2018).

40. Frydendahl, C. et al. Optical reconfiguration and polarization control in semi-continuous gold films close to the percolation threshold. *Nanoscale* **9**, 12014-12024 (2017).

41. Yoon, Y., Ganapathi, K. & Salahuddin, S. How Good Can Monolayer $MoS_2$ Transistors Be? *Nano Letters* **11**, 3768-3773 (2011).

42. Liu, W., Sarkar, D., Kang, J.H., Cao, W. & Banerjee, K. Impact of Contact on the Operation and Performance of Back-Gated Monolayer MoS2 Field-Effect-Transistors. *ACS Nano* **9**, 7904-7912 (2015).

43. Kappera, R. et al. Phase-engineered low-resistance contacts for ultrathin $MoS_2$ transistors. *Nature Materials* **13**, 1128-1134 (2014).
44. Qi, Y.P. et al. Superconductivity in Weyl semimetal candidate $MoTe_2$. *Nature Communications* **7**, 11038 (2016).
45. Xi, X. et al. Ising pairing in superconducting NbSe2 atomic layers. *Nature Physics* **12**, 139-143 (2016).
46. Blochl, P.E. Projector Augmented-Wave Method. *Physical Review B* **50**, 17953-17979 (1994).
47. Kresse, G. & Joubert, D. From ultrasoft pseudopotentials to the projector augmented-wave method. *Physical Review B* **59**, 1758-1775 (1999).
48. Kresse, G. & Furthmuller, J. Efficient iterative schemes for ab initio total-energy calculations using a plane-wave basis set. *Physical Review B* **54**, 11169-11186 (1996).
49. Klimes, J., Bowler, D.R. & Michaelides, A. Van der Waals density functionals applied to solids. *Physical Review B* **83**, 195131 (2011).
50. Ferrari, A.C. et al. Raman spectrum of graphene and graphene layers. *Physical Review Letters* **97**, 187401 (2006).
51. Lee, C. et al. Anomalous Lattice Vibrations of Single- and Few-Layer $MoS_2$. *ACS Nano* **4**, 2695-2700 (2010).
52. Zhang, H. et al. Interference effect on optical signals of monolayer $MoS_2$. *Applied Physics Letters* **107**, 101904 (2015).
53. Du, L.J. et al. 2D proximate quantum spin liquid state in atomic-thin alpha-$RuCl_3$. *2D Materials* **6**, 015014 (2019).
54. Zhou, B.Y. et al. Possible structural transformation and enhanced magnetic fluctuations in exfoliated alpha-$RuCl_3$. *Journal of Physics and Chemistry of Solids* **128**, 291-295 (2019).
55. Shi, J.P. et al. Monolayer MoS2 Growth on Au Foils and On-Site Domain Boundary Imaging. *Advanced Functional Materials* **25**, 842-849 (2015).
56. Liu, G.D. et al. Development of a vacuum ultraviolet laser-based angle-resolved photoemission system with a superhigh energy resolution better than 1 meV. *Review of Scientific Instruments* **79**, 023105 (2008).

# Supplementary Information for

## Universal mechanical exfoliation of large-area 2D crystals

Yuan Huang[†], Yu-Hao Pan[†], Rong Yang[†], Li-Hong Bao, Lei Meng, Hai-Lan Luo, Yong-Qing Cai, Guo-Dong Liu, Wen-Juan Zhao, Zhang Zhou, Liang-Mei Wu, Zhi-Li Zhu, Ming Huang, Li-Wei Liu, Lei Liu, Peng Cheng, Ke-Hui Wu, Shi-Bing Tian, Chang-Zhi Gu, You-Guo Shi, Yan-Feng Guo, Zhi Gang Cheng, Jiang-Ping Hu, Lin Zhao, Eli Sutter, Peter Sutter[*], Ye-Liang Wang, Wei Ji[*], Xing-Jiang Zhou[*], and Hong-Jun Gao[*]

[†]These authors contributed equally to this work.

*Correspondence to: psutter@unl.edu (P.S.); wji@ruc.edu.cn (W.J.); xjzhou@iphy.ac.cn (X.J.Z.); hjgao@iphy.ac.cn (H.J.G.)

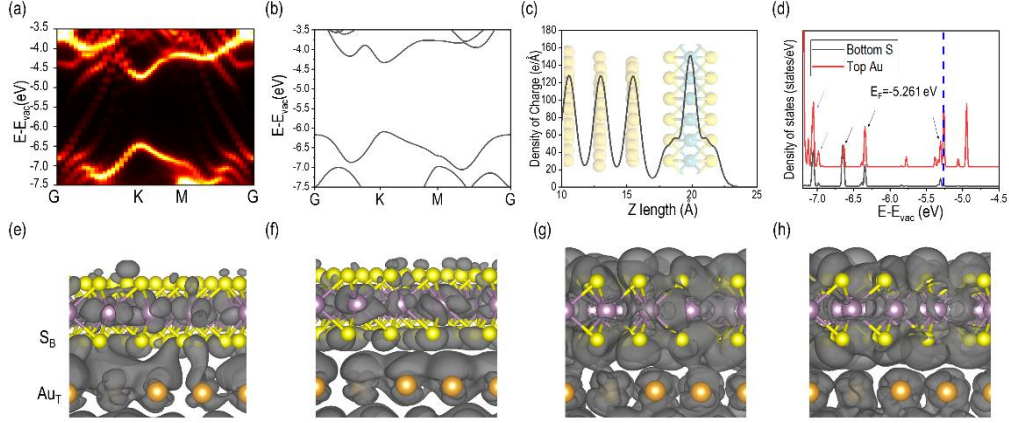

**Fig. S1. Electronic structures of an Au (111) supported monolayer MoS$_2$ (1L-MoS$_2$), a typical 2D material, interface.** Here, MoS$_2$ was chosen in consideration of that it has a moderate $R_{LA/IL}$ of 1.51 and group 16 (VIA) elements are the major portion of non-metal elements in all considered 2D layers. The vacuum energy was set as energy zero in both bandstructures and DOS plots. **(a)** Unfolded bandstructure of the interface, which was calculated using the KPROJ program based on the k-projection method[1,2]. **(b)** Bandstructure of freestanding 1L-MoS$_2$. The bandgap of 1.76 eV and the shape of valence (VB) and conduction (CB) bands remain nearly unchanged after the free-standing layer in contact with the Au substrate while the positions of VB and CB shift downward by ~0.5 eV. This shift is a result of vertical electric dipole moments (~0.75 $e·Å$) formed at the interface upon contact, which were illustrated by the DCD in Fig. 1e and a layer-averaged line profile associated with total electronic charge density, as plotted in **(c)**. Dipole moment $p$ here is defined as $p = -e\sum_l Z_l R_l + \int dr\, r\rho(\mathbf{r})$, where $e$ is the electron charge, the $l$ summation is over Mo and S ionic sites, and $\rho(\mathbf{r})$ is the electronic charge density in the MoS$_2$ interlayer. Both VB and CB of the intact MoS$_2$ were perturbed by some Au states, as depicted in **(a)**, which indicates Au states electronically hybridize with states of MoS$_2$. We plotted local density of states (LDOS) of the interface in **(d)** in order to reveal the interactions between MoS$_2$ and Au (111). Only the Gamma point was used for clarity. Grey and red lines represent the LDOSs of an interfacial S atom (denote $S_B$) and an interfacial Au atom (denote $Au_T$). The hybrid states of $S_B$ and $Au_T$ were marked by black and light blue arrows. Each of those peaks contains six states while some of those states were representatively visualized in **(e)** to **(h)**. Panels **(e)** and **(f)** plot the isosurface contours of wavefunction norms for the bonding and anti-bonding states sitting at ~ -5.3 eV with an isosurface value of $1\times10^{-4}$ $e$/Bohr$^3$, while those for the states at ~ -7.0 eV were shown in **(g)** and **(h)** with an isosurface value of $3\times10^{-5}$ $e$/Bohr$^3$, respectively. The hybridization between them, however, is not as strong as a covalent bond. The energy splitting of bonding-states and anti-bonding states is only few meV for those states marked by the black arrows and is up to 86 meV for those states marked by the light-blue arrows. While both the bonding and anti-bonding states all fully occupied, together with the rather small energy splitting, we conclude the interaction between Au and MoS$_2$ is not covalent bonding, but an interaction type

called covalent-like quasi-bonding (CLQB), as we discussed in the main text and recently identified in 2D layers.

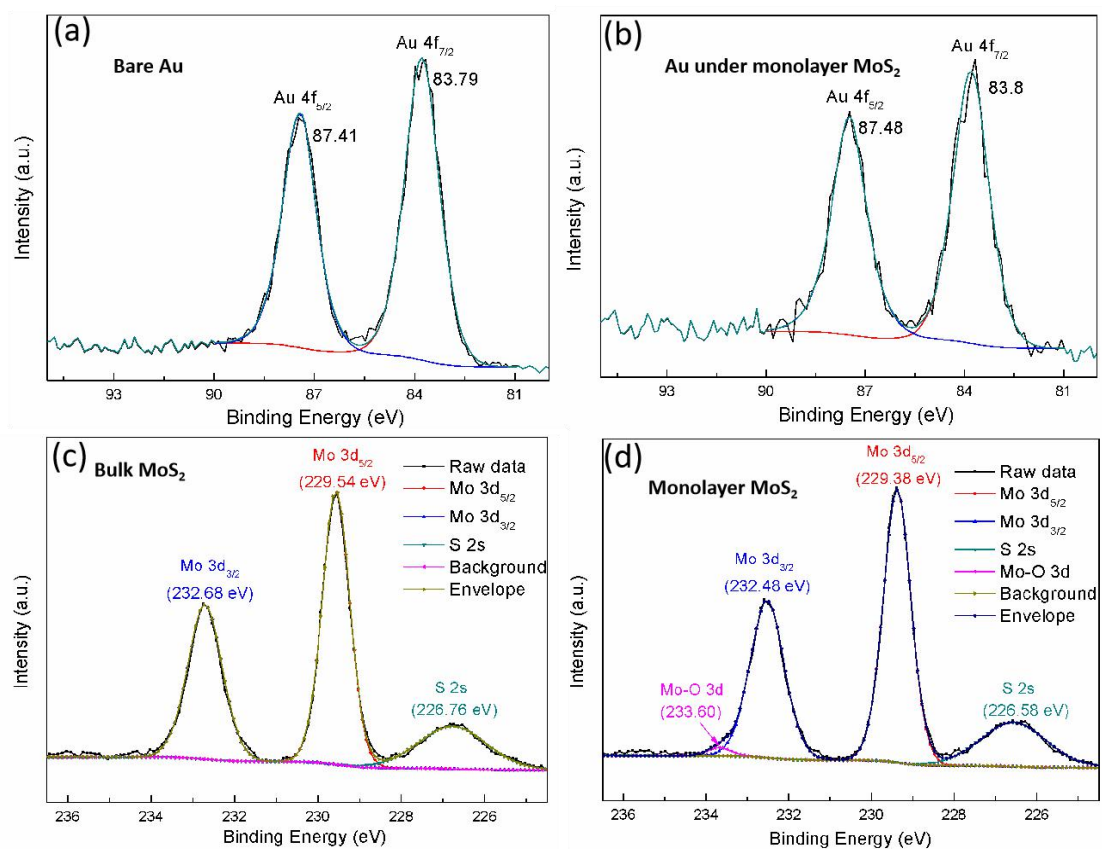

**Fig. S2. (a, b)** XPS spectra of Au 4f from bare Au surface and the Au beneath a monolayer $MoS_2$ flake. **(c, d)** Mo 3d and S 2s spectra and the corresponding fitting curves of a bulk $MoS_2$ flake and of an exfoliated single layer $MoS_2$ flake.

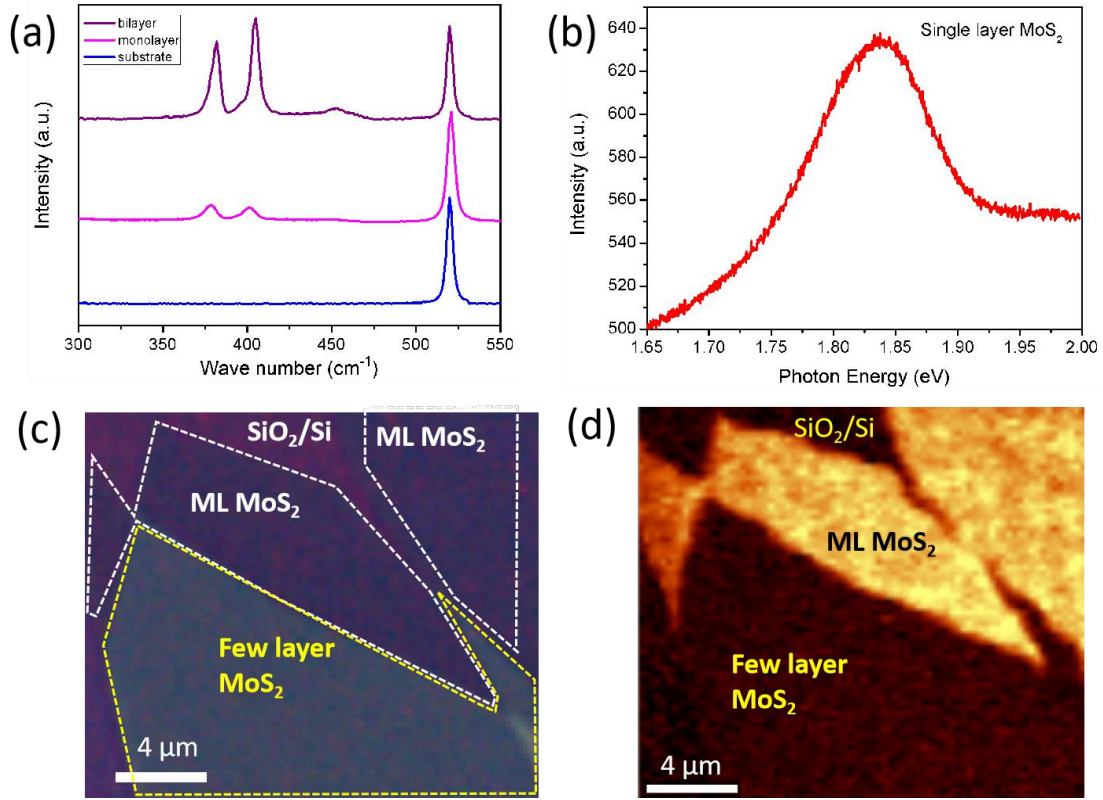

**Fig. S3**. **Raman and PL spectra of MoS$_2$ exfoliated with Au/Ti adhesion layer (nominally 1.5 nm/0.5 nm).** **(a)** Raman spectra of monolayer and bilayer MoS$_2$, normalized by the intensity of the Si peak at 520 cm$^{-1}$. **(b)** PL spectrum of monolayer MoS$_2$ exfoliated onto the ultrathin Au/Ti adhesion layer. **(c)** Optical image of a MoS$_2$ flake with coexisting monolayer and few-layer areas. **(d)** PL mapping image of the MoS$_2$ flake shown in **(c)**, with PL intensity in the monolayer area substantially higher than in few-layer and substrate regions.

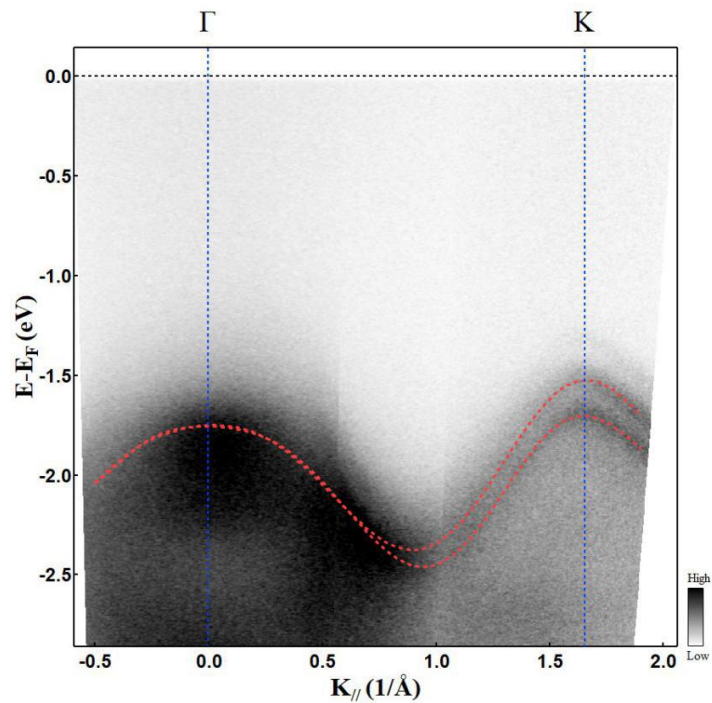

**Fig. S4.** Band structure of monolayer MoS$_2$ flake exfoliated on Au/Ti adhesion layer (5 nm/1 nm).

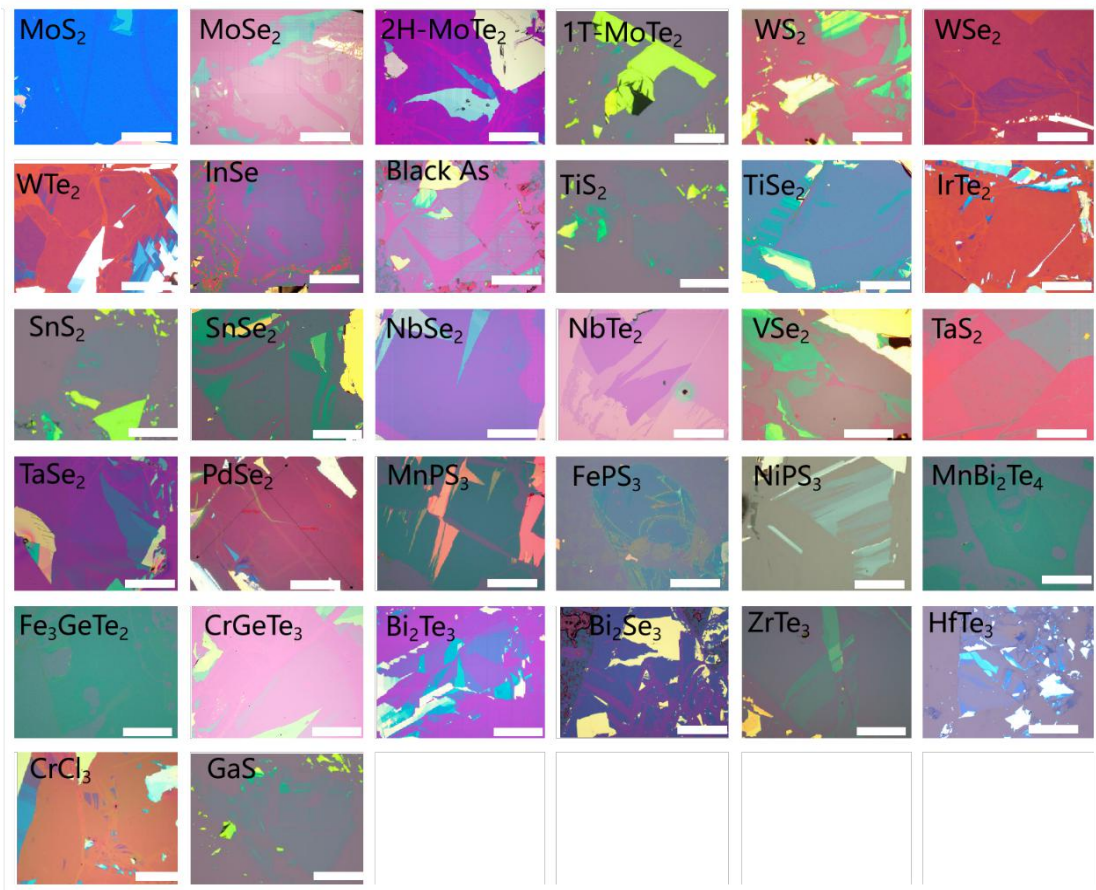

**Fig. S5.** Optical images of 2D materials exfoliated using Au adhesion layers. Some of materials shown are obtained here for the first time in monolayer form, for example, IrTe$_2$, ZrTe$_3$, PtSe$_2$. Scale bars: 500 μm.

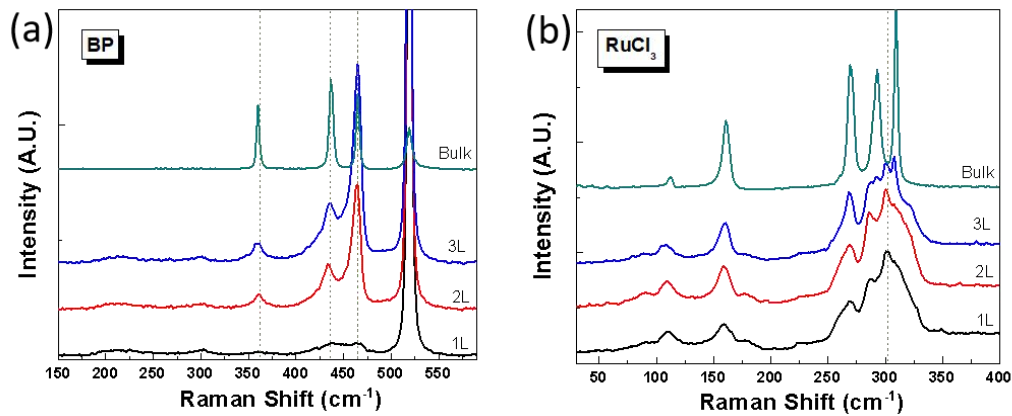

**Fig. S6.** Raman spectra of exfoliated black phosphorus (BP) and RuCl$_3$ crystals with different thickness, in comparison with the corresponding layered bulk crystals.

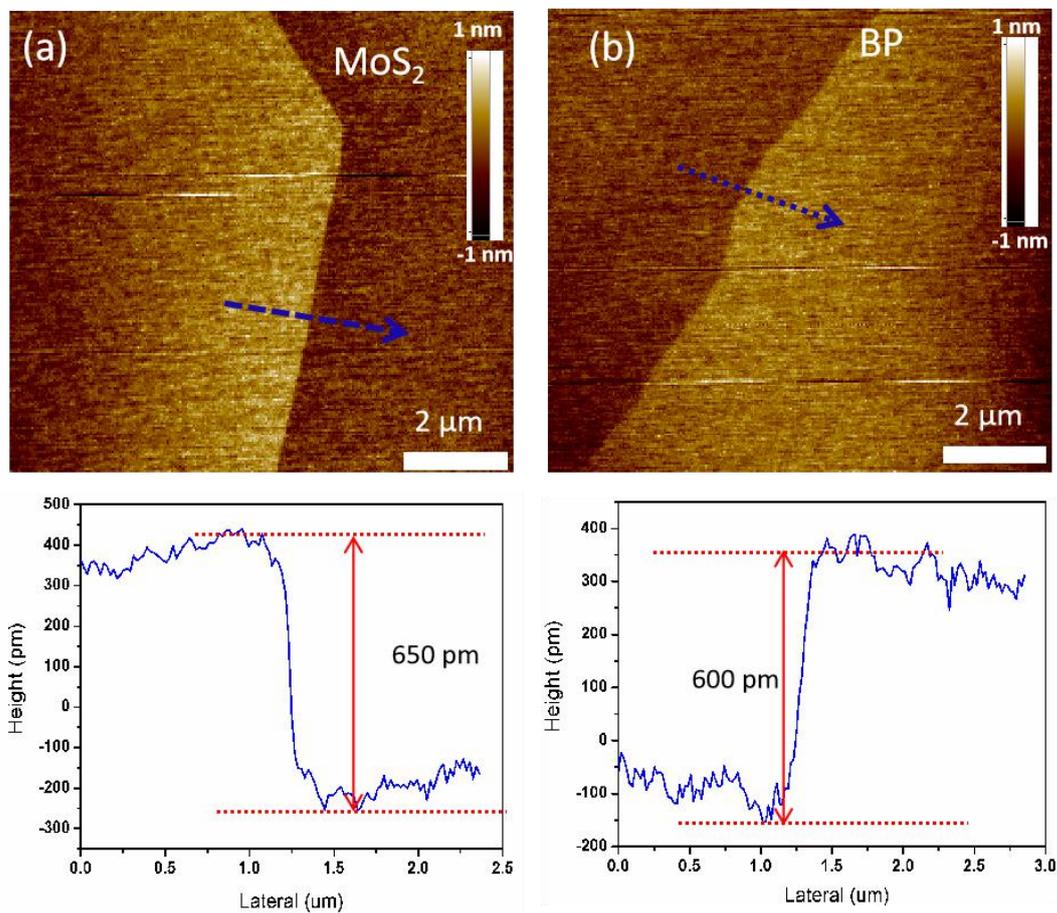

**Fig. S7**. AFM images of as-exfoliated 2D MoS$_2$ and black phosphorus/phosphorene (BP).

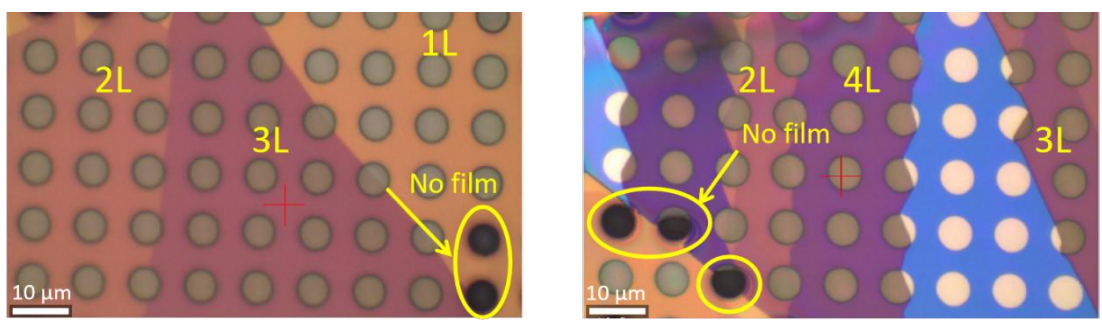

**Fig. S8**. Suspended WSe$_2$ samples with suspended coverage of 97% and 93%, respectively.

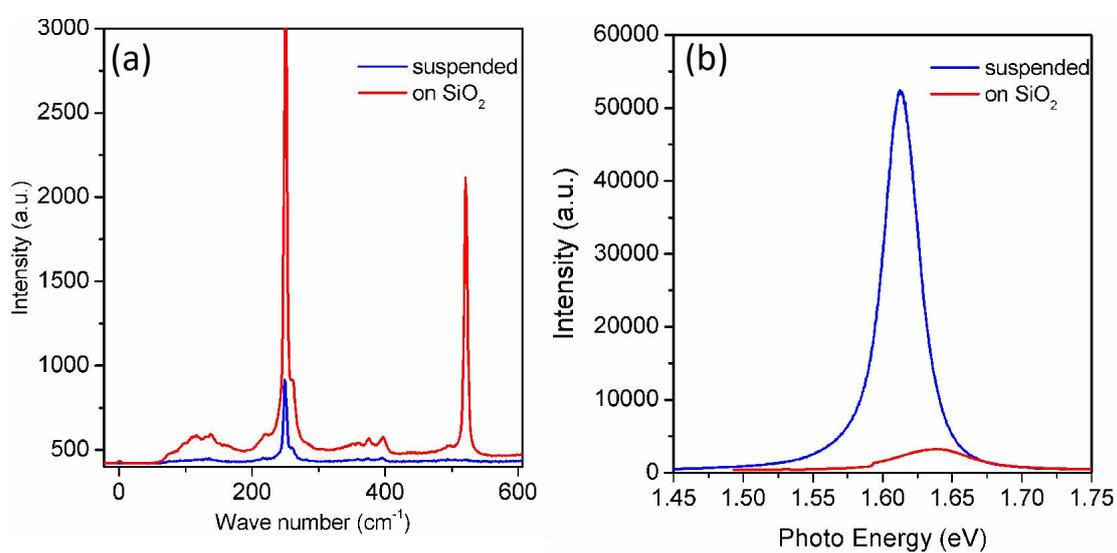

**Fig. S9.** Raman and PL spectra of suspended monolayer WSe$_2$ in comparison with monolayer WSe$_2$ supported on a SiO$_2$/Si substrate.

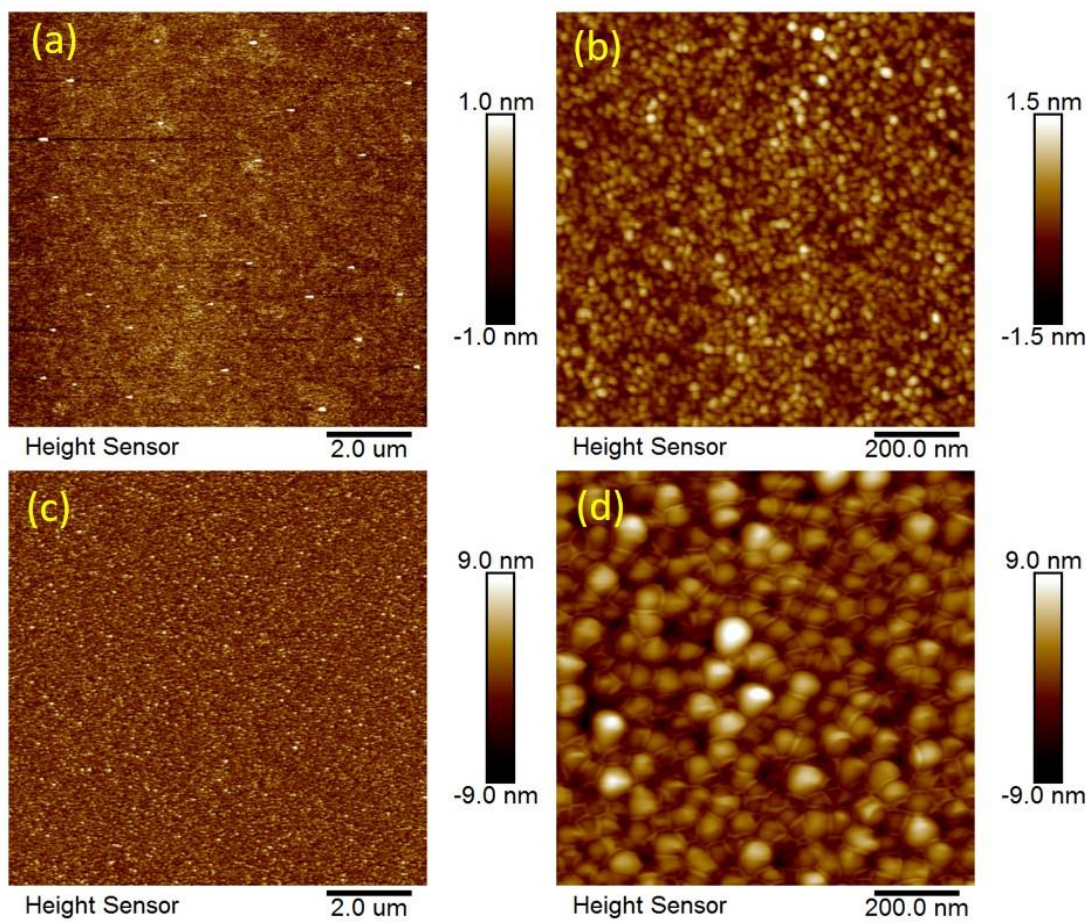

**Fig. S10.** AFM images of Au/Ti (1.5 nm/ 0.5 nm) adhesion layers before and after annealing (250 $^0$C, 2h).

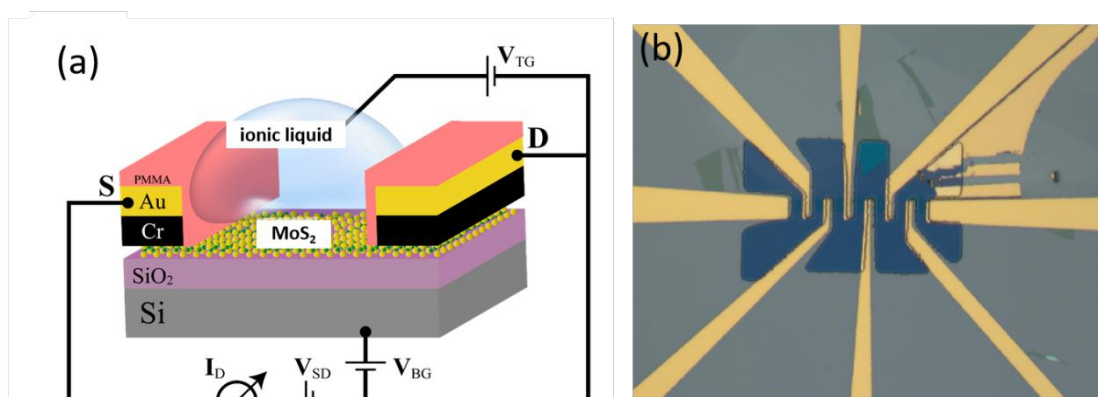

**Fig. S11. (a)** Monolayer field-effect transistor (FET) device structure with ionic liquid top gate. **(b)** Optical micrograph of an actual FET device with monolayer MoS$_2$ channel, and an insulating PMMA coating with window for the ionic liquid top gate.

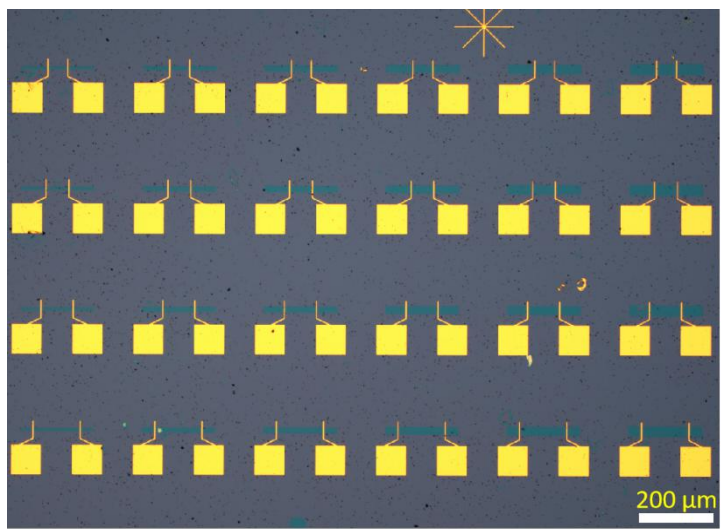

**Fig. S12.** Optical image of UV-patterned MoS$_2$ device arrays fabricated from a single large monolayer MoS$_2$ flake exfoliated onto an ultrathin Au/Ti adhesion layer.

# Table S1. Calculated energies of all 58 considered 2D materials

| 2D-Materials | Ads. Energy on Au(111) (eV)/unit cell | Interlayer Coupling Energy (eV)/unit cell | Ads. Energy on Au(111) (eV)/Å² | Interlayer Coupling Energy (eV)/Å² | Ads. Energy on Au(111) (eV)/Bottom Atom | Interlayer Coupling Energy (eV)/Bottom Atom | $R_{LA/IL}$ | Space Group | Magnetic Structure* |
|---|---|---|---|---|---|---|---|---|---|
| h-BN | 0.146 | 0.136 | 0.0268 | 0.0250 | 0.073 | 0.068 | 1.07 | P -6 m 2 | NM |
| Gr | 0.147 | 0.118 | 0.0279 | 0.0225 | 0.073 | 0.059 | 1.24 | P 63/m m c | NM |
| P(Black) | 0.801 | 0.484 | 0.0556 | 0.0334 | 0.400 | 0.242 | 1.67 | C m c a | NM |
| As(Black) | 1.014 | 0.584 | 0.0597 | 0.0350 | 0.507 | 0.292 | 1.71 | C m c a | NM |
| $W_2N_3$ | 0.520 | 0.230 | 0.0716 | 0.0316 | 0.520 | 0.230 | 2.26 | P 63/m m c | NM |
| $TiS_2$ | 1.087 | 0.542 | 0.0547 | 0.0273 | 0.543 | 0.271 | 2.00 | P -3 m 1 | NM |
| $VS_2$ | 0.412 | 0.265 | 0.0476 | 0.0305 | 0.412 | 0.265 | 1.56 | P -3 m 1 | FM |
| $NbS_2$ | 0.605 | 0.264 | 0.0632 | 0.0276 | 0.605 | 0.264 | 2.29 | R 3 m H | NM |
| $TaS_2$-2H | 0.549 | 0.249 | 0.0575 | 0.0262 | 0.549 | 0.249 | 2.19 | P 63/m m c | NM |
| $TaS_2$-1T | 0.457 | 0.278 | 0.0469 | 0.0286 | 0.457 | 0.278 | 1.64 | P -3 m 1 | NM |
| $MoS_2$ | 0.345 | 0.229 | 0.0397 | 0.0263 | 0.345 | 0.229 | 1.51 | P 63/m m c | NM |
| $WS_2$ | 0.329 | 0.227 | 0.0371 | 0.0261 | 0.329 | 0.227 | 1.42 | P 63/m m c | NM |
| $ReS_2$ | 1.128 | 0.856 | 0.0312 | 0.0211 | 0.282 | 0.214 | 1.48 | P -1 | NM |
| $CoPS_3$ | 4.883 | 1.433 | 0.0829 | 0.0211 | 1.221 | 0.358 | 3.35 | C1 2/m 1 | FM |
| $NiPS_3$ | 1.946 | 1.354 | 0.0336 | 0.0233 | 0.486 | 0.339 | 1.44 | C1 2/m 1 | FM |
| $PdS_2$ | 1.479 | 0.889 | 0.0494 | 0.0290 | 0.739 | 0.445 | 1.71 | P b c a | NM |
| $PtS_2$ | 0.457 | 0.292 | 0.0421 | 0.0262 | 0.457 | 0.292 | 1.60 | P -3 m 1 | NM |
| GaS | 0.388 | 0.218 | 0.0347 | 0.0196 | 0.194 | 0.109 | 1.78 | P 63/m m c | NM |
| InS | 1.093 | 1.025 | 0.0718 | 0.0588 | 0.547 | 0.512 | 1.22 | P m n n | NM |
| $GeS_2$ | 0.316 | 0.273 | 0.0262 | 0.0223 | 0.316 | 0.273 | 1.17 | P 42/n m c Z | NM |
| SnS | 0.90 | 0.617 | 0.0534 | 0.0365 | 0.451 | 0.309 | 1.46 | P n m a | NM |
| $SnS_2$ | 0.488 | 0.243 | 0.0418 | 0.0207 | 0.488 | 0.243 | 2.02 | P -3 m 1 | NM |
| $TiSe_2$ | 0.483 | 0.289 | 0.0454 | 0.0270 | 0.483 | 0.289 | 1.68 | P -3 m 1 | NM |
| $VSe_2$ | 0.476 | 0.261 | 0.0502 | 0.0272 | 0.476 | 0.261 | 1.85 | P -3 m 1 | FM |
| $NbSe_2$ | 0.607 | 0.304 | 0.0590 | 0.0295 | 0.607 | 0.304 | 2.00 | P 63/m m c | NM |
| $MoSe_2$ | 0.414 | 0.243 | 0.0441 | 0.0258 | 0.414 | 0.243 | 1.71 | P 63/m m c | NM |
| $WSe_2$ | 0.384 | 0.241 | 0.0409 | 0.0256 | 0.384 | 0.241 | 1.60 | P 63/m m c | NM |
| $ReSe_2$ | 1.458 | 0.906 | 0.0373 | 0.0232 | 0.364 | 0.227 | 1.61 | P -1 | NM |
| FeSe | 0.565 | 0.381 | 0.0424 | 0.0285 | 0.565 | 0.381 | 1.49 | C m m a | AFM |
| $PdSe_2$ | 1.284 | 1.117 | 0.0388 | 0.0288 | 0.642 | 0.559 | 1.34 | P b c a | NM |
| $PtSe_2$ | 0.641 | 0.341 | 0.0537 | 0.0276 | 0.641 | 0.341 | 1.95 | P -3 m 1 | NM |
| GaSe | 0.352 | 0.228 | 0.0285 | 0.0185 | 0.176 | 0.114 | 1.54 | P 63/m m c | NM |
| InSe | 0.519 | 0.267 | 0.0369 | 0.0188 | 0.259 | 0.133 | 1.97 | P 63/m m c | NM |
| $GeSe_2$ | 0.411 | 0.335 | 0.0304 | 0.0242 | 0.411 | 0.335 | 1.25 | P 42/n m c Z | NM |
| SnSe | 1.033 | 0.658 | 0.0560 | 0.0353 | 0.516 | 0.329 | 1.58 | P n m a | NM |
| $SnSe_2$ | 0.621 | 0.275 | 0.0488 | 0.0214 | 0.621 | 0.275 | 2.24 | P -3 m 1 | NM |
| $Sb_2Se_3$ | 0.630 | 0.340 | 0.0448 | 0.0239 | 0.630 | 0.340 | 1.88 | R -3 m H | NM |
| $Bi_2Se_3$ | 0.618 | 0.336 | 0.0416 | 0.0224 | 0.618 | 0.336 | 1.86 | R -3 m H | NM |
| $ZrTe_3$ | 1.283 | 0.633 | 0.0554 | 0.0274 | 0.428 | 0.211 | 2.02 | P 1 21/m 1 | NM |
| $ZrTe_5$ | 2.378 | 1.088 | 0.0431 | 0.0186 | 1.189 | 0.544 | 2.32 | C m c m | NM |
| $HfTe_3$ | 1.307 | 0.637 | 0.0567 | 0.0279 | 0.436 | 0.212 | 2.03 | P 1 21/m 1 | NM |
| $HfTe_5$ | 1.538 | 1.078 | 0.0280 | 0.0186 | 0.308 | 0.216 | 1.51 | C m c m | NM |
| $VTe_2$ | 0.658 | 0.295 | 0.0585 | 0.0262 | 0.658 | 0.295 | 2.23 | P -3 m 1 | FM |
| $MoTe_2$ | 0.490 | 0.281 | 0.0456 | 0.0261 | 0.490 | 0.281 | 1.75 | P 63/m m c | NM |
| $WTe_2$ | 1.026 | 0.516 | 0.0471 | 0.0235 | 0.513 | 0.258 | 2.00 | P m n 21 | NM |
| $MnBi_2Te_4$ | 0.791 | 0.368 | 0.0497 | 0.0229 | 0.791 | 0.368 | 2.17 | R -3 m H | FM |
| FeTe | 0.654 | 0.329 | 0.0461 | 0.0232 | 0.654 | 0.329 | 1.99 | C m m a | AFM |
| $IrTe_2$ | 0.870 | 0.683 | 0.0680 | 0.0503 | 0.870 | 0.683 | 1.35 | P -3 m 1 | NM |
| $PtTe_2$ | 0.865 | 0.479 | 0.0631 | 0.0334 | 0.865 | 0.479 | 1.89 | P -3 m 1 | NM |
| $Sb_2Te_3$ | 0.773 | 0.404 | 0.0489 | 0.0256 | 0.773 | 0.404 | 1.91 | R -3 m H | NM |
| $Bi_2Te_3$ | 0.842 | 0.397 | 0.0508 | 0.0235 | 0.842 | 0.397 | 2.16 | R -3 m H | NM |
| $CrCl_3$ | 0.800 | 0.550 | 0.0300 | 0.0180 | 0.267 | 0.183 | 1.67 | R -3 H | FM |
| $RuCl_3$ | 1.111 | 0.574 | 0.0356 | 0.0184 | 0.370 | 0.191 | 1.94 | P 31 1 2 | AFM |
| $CdCl_2$ | 0.260 | 0.203 | 0.0204 | 0.0158 | 0.260 | 0.203 | 1.29 | R -3 m H | NM |
| $CrBr_3$ | 1.693 | 1.207 | 0.0245 | 0.0175 | 0.282 | 0.201 | 1.40 | R -3 H | FM |
| $CdBr_2$ | 0.323 | 0.222 | 0.0234 | 0.0160 | 0.323 | 0.222 | 1.47 | R -3 m H | NM |
| $CrI_3$ | 1.425 | 0.715 | 0.0344 | 0.0175 | 0.475 | 0.119 | 1.96 | R -3 H | FM |
| $CdI_2$ | 0.387 | 0.247 | 0.0248 | 0.0157 | 0.387 | 0.247 | 1.57 | R -3 m H | NM |

*NM: Non-magnetic; FM: Ferromagnetic; AFM: Antiferromagnetic

**Reference:**

1   Mingxing Chen* and M. Weinert, Layer k-projection and unfolding electronic bands at interfaces, Phys. Rev. B 98, 245421 (2018).
2   M. X. Chen*, W. Chen, Zhenyu Zhang and M. Weinert, Effects of magnetic dopants in $(Li_{0.8}M_{0.2}OH)FeSe$ (M = Fe, Mn, Co): Density functional theory study using a band unfolding technique, Phys. Rev. B 96, 245111 (2017).